\RequirePackage[2020-02-02]{latexrelease}

\documentclass[%
 reprint,
superscriptaddress,
 amsmath,amssymb,
 aps,
 prl,
]{revtex4-2}

\usepackage{graphicx}
\usepackage{dcolumn}
\usepackage{bm}
\usepackage{hyperref}
\usepackage{xcolor}
\usepackage{booktabs}
\usepackage[normalem]{ulem}

\begin{document}

\title{High-precision mass measurement of $^{24}$Si and a refined determination of the $rp$ process at the $A=22$ waiting point}


\author{D. Puentes}
\email[]{puentes@frib.msu.edu}
\affiliation{Department of Physics and Astronomy, Michigan State University, East Lansing, Michigan 48824, USA}
\affiliation{National Superconducting Cyclotron Laboratory, East Lansing, Michigan 48824, USA}
\author{Z. Meisel}
\affiliation{Institute of Nuclear \& Particle Physics, Department of Physics and Astronomy, Ohio University, Athens, Ohio 45701, USA}
\affiliation{Joint Institute for Nuclear Astrophysics -- Center for the Evolution of the Elements, Michigan State University, East Lansing, Michigan 48824, USA}
\author{G. Bollen}
\affiliation{Department of Physics and Astronomy, Michigan State University, East Lansing, Michigan 48824, USA}
\affiliation{Facility for Rare Isotope Beams, East Lansing, Michigan 48824, USA}
\author{A. Hamaker}
\affiliation{Department of Physics and Astronomy, Michigan State University, East Lansing, Michigan 48824, USA}
\affiliation{National Superconducting Cyclotron Laboratory, East Lansing, Michigan 48824, USA}
\author{C. Langer}
\affiliation{University of Applied Sciences Aachen, 52066 Aachen, Germany}
\author{E. Leistenschneider}
\affiliation{National Superconducting Cyclotron Laboratory, East Lansing, Michigan 48824, USA}
\author{C. Nicoloff}
\affiliation{Department of Physics and Astronomy, Michigan State University, East Lansing, Michigan 48824, USA}
\affiliation{National Superconducting Cyclotron Laboratory, East Lansing, Michigan 48824, USA}
\author{W.-J. Ong}
\affiliation{Nuclear and Chemical Sciences Division, Lawrence Livermore National Laboratory, Livermore, California 94550, USA}
\affiliation{Joint Institute for Nuclear Astrophysics -- Center for the Evolution of the Elements, Michigan State University, East Lansing, Michigan 48824, USA}
\author{M. Redshaw}
\affiliation{National Superconducting Cyclotron Laboratory, East Lansing, Michigan 48824, USA}
\affiliation{Department of Physics, Central Michigan University, Mount Pleasant, Michigan 48859, USA}
\author{R. Ringle}
\affiliation{National Superconducting Cyclotron Laboratory, East Lansing, Michigan 48824, USA}
\author{C. S. Sumithrarachchi}
\affiliation{National Superconducting Cyclotron Laboratory, East Lansing, Michigan 48824, USA}
\author{J. Surbrook}
\affiliation{Department of Physics and Astronomy, Michigan State University, East Lansing, Michigan 48824, USA}
\affiliation{National Superconducting Cyclotron Laboratory, East Lansing, Michigan 48824, USA}
\author{A. A. Valverde}
\affiliation{Department of Physics and Astronomy, University of Manitoba, Winnipeg, MB R3T 2N2, Canada}
\author{I. T. Yandow}
\affiliation{Department of Physics and Astronomy, Michigan State University, East Lansing, Michigan 48824, USA}
\affiliation{National Superconducting Cyclotron Laboratory, East Lansing, Michigan 48824, USA}

\date{\today}

\begin{abstract}
We report a high precision mass measurement of $^{24}{\rm Si}$, performed with the LEBIT facility at the National Superconducting Cyclotron Laboratory. The atomic mass excess, $10\;753.8$(37) keV, is a factor of 5 more precise than previous results. This substantially reduces the uncertainty of the $^{23}{\rm Al}(p,\gamma)^{24}{\rm Si}$ reaction rate, which is a key part of the rapid proton capture ($rp$) process powering Type I X-ray bursts. The updated rate constrains the onset temperature of the $(\alpha,p)$ process at the $^{22}{\rm Mg}$ waiting-point to a precision of 9\%.
\end{abstract}

\maketitle


Type I X-ray bursts occur at astrophysical sites consisting of a binary star system where a neutron star accretes H/He-rich matter from a companion star, leading to nuclear burning on the neutron star surface~\cite{Lewin1993, Schatz2006}. As the temperature increases, the hot CNO cycle proceeds until the triple-$\alpha$ process is initiated with an ensuing thermonuclear runaway. The thermonuclear runaway explosion consists of multiple competing reaction sequences including the rapid proton ($rp$)~\cite{Schatz1998} and ($\alpha, p$)~\cite{Wallace1981} processes. The rate at which the runaway occurs as well as the reaction sequence competition depends on factors such as accretion rate and fuel composition. The only observable from these explosions, however, is the X-ray burst light curve, which can be recorded with telescopes like the \textit{Neutron Star Interior Composition Explorer} (NICER)~\cite{Bult2021}. 

In addition to the astrophysical conditions, the main determinant of the shape of the light curve is the nuclear physics involved~\cite{Meisel2018}. The nuclear physics input consists of the fast ($p,\gamma$), ($\alpha, \gamma$) and ($\alpha, p$) reaction rates and slow $\beta^+$ decays. Single-zone~\cite{Schatz2001} and multi-zone~\cite{Heger2007,Zamfir2012} X-ray burst models have been used to understand the impact of the uncertainties derived from these nuclear physics inputs \cite{Cyburt2016, Parikh2008}. Through these analyses, it was shown that only a few reactions play a large role in the uncertainties of the simulated light curves~\cite{Cyburt2016}. The dramatic variations in the simulated light curve due to these reaction rate uncertainties can make it difficult to extract astrophysical parameters from comparisons with observed bursting events.

One of those influential reactions is the $^{23}$Al($p,\gamma$)$^{24}$Si reaction. Variations within the uncertainty of this reaction rate lead to significant shifts in the simulated X-ray light curve. The importance of this reaction stems from its direct connection to the $^{22}$Mg waiting point along the $rp$ process~\cite{Fisker2004}. The $Q$ value for $^{22}{\rm Mg}(p,\gamma)$ (0.141 MeV)~\cite{Huan2021} is small enough to establish a ($\gamma,p$)-($p,\gamma$) equilibrium~\cite{VanWormer1994}, which makes it difficult for the $rp$ process to proceed in this mass region. There is evidence~\cite{VanWormer1994} that the $^{22}$Mg waiting point can be bypassed via the $^{22}$Mg($\alpha,p$) reaction. However, it is also possible for the $^{23}{\rm Al}(p,\gamma)$ reaction to undergo resonant capture to only a few low lying states in $^{24}$Si when certain conditions are assumed for X-ray bursts. This network competition in the reaction flow has been investigated in previous experiments, providing constraints on when $^{22}$Mg($\alpha,p$) becomes significant during X-ray burst events~\cite{Randhawa2020}. However, uncertainties in the $^{23}$Al($p,\gamma$) reaction need to be resolved in order to confirm the flow of the $rp$ process at this waiting point. A recent experiment measured $\gamma$ rays from $^{23}{\rm Al}(d,n)$~\cite{Wolf2019} to identify unbound states in $^{24}$Si, making the uncertainty in the mass of $^{24}$Si (19 keV ~\cite{Huan2021}) the dominant source of uncertainty for the $^{23}{\rm Al}(p,\gamma)$ reaction rate. A high-precision mass measurement of $^{24}$Si would constrain the $rp$ process in this mass region by precisely quantifying the bypass of the $^{22}{\rm Mg}$ waiting point. In this Letter, we present a new high-precision mass measurement of $^{24}$Si using Penning trap mass spectrometry.

A beam of $^{24}$Si was produced at the National Superconducting Cyclotron Laboratory (NSCL)~\cite{Morrissey2003} via projectile fragmentation of a $^{28}$Si$^{14+}$ beam, at an energy of 160 A MeV, with a 1530 mg/cm$^2$ thin beryllium foil target. The resulting cocktail beam was purified using the A1900 Fragment Separator~\cite{Morrissey2003}. Afterwards, the beam was sent through a momentum compression beamline, passing through aluminum degraders and an Al wedge to degrade the beam energy before entering the gas stopper~\cite{Sumithrarachchi2020}. Interactions between the $^{24}$Si ions and helium gas in the stopper slowed the rare isotope ions, which were then extracted using a combination of direct current and radio frequency (RF) fields. After extraction from the gas cell, the rare isotope ions were injected into a Radio-Frequency Quadrupole (RFQ), accelerated to an energy of 30 keV and purified with a magnetic dipole mass separator based on their mass-to-charge ratio (A/Q). Following the beam stopping facility, $^{24}$Si was delivered as a molecular ion, [$^{24}$SiO$_2$H]$^+$ at $A=57$. The beam was then transported to the Low Energy Beam and Ion Trap (LEBIT)~\cite{Ringle2013} facility where the mass measurements were performed. 

The first component of the LEBIT facility is a beam cooler and buncher \cite{Schwartz2016}. The ions were cooled, collected, and ejected as a pulsed beam. The pulsed ion beam was subsequently injected into the 9.4 T Penning trap mass spectrometer~\cite{Ringle2009} and directed off-axis relative to the trap center using Lorentz steerers \cite{Ringle2007}, capturing the ions in an initial magnetron orbit. During trapping, contaminants were purged using a sum of dipole RF excitations at the identified contaminants' reduced cyclotron frequencies and a broadband SWIFT excitation~\cite{Blaum2004,Kwiatkowski2015} to remove contaminant ions. The major contaminants identified during the experiment were [C$_3$H$_2$F]$^+$, [C$_2$HS]$^+$ and [C$_3$H$_5$O]$^+$. The Time-of-Flight Ion Cyclotron Resonance (TOF-ICR) technique was then used to measure the mass of the ion of interest. In this method, a quadrupole excitation is performed for a range of frequencies near the cyclotron frequency, $\nu_c = qB/2\pi m$, of the ion of interest~\cite{Bollen1990,Konig1995}. As the RF excitation, $\nu_{RF}$, approaches $\nu_c$, the magnetron motion is converted to reduced cyclotron motion. The conversion is maximized when $\nu_{RF} \approx \nu_c$. Following the excitation, the ions are ejected from the trap to a microchannel plate (MCP) detector, where the time of flight is recorded. During the conversion, the ions gain energy, which is observed as a minimum in the TOF response curve. The theoretical line shape can be fit to the data to extract $\nu_c$ \cite{Konig1995}.

\begin{figure}
\centering
\includegraphics[scale=0.33]{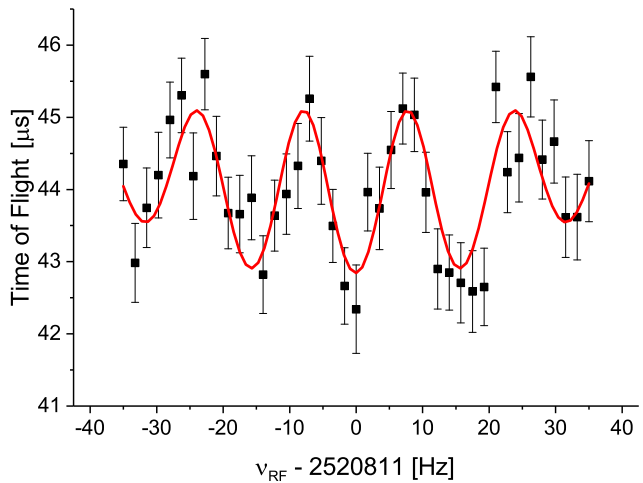}
\caption{An example of a 75-ms [$^{24}$SiO$_2$H]$^+$ time-of-flight Ramsey cyclotron resonance curve from which $\nu_c$ was determined. The line shape is described in Ref.  \cite{Kretzschmar2007}.}
\label{fig:Resonance}
\end{figure}

Another excitation scheme, known as a Ramsey excitation~\cite{Kretzschmar2007}, was also performed in this experiment. This generates a TOF pattern visible in FIG. \ref{fig:Resonance} and described by Ref~\cite{Kretzschmar2007}, which allows superior precision to the simple RF excitation scheme for the same ion trapping times. 18 measurements of [$^{24}$SiO$_2$H]$^+$ were performed, consisting of 6 normal quadrupole excitation measurements and 12 Ramsey excitation measurements. The normal quadrupole excitation measurements were performed with 50 ms and 75 ms excitations. The Ramsey measurements were performed with 25 ms, 50 ms, 75 ms, 100 ms and 150 ms excitations.

Cyclotron frequency measurements of [$^{24}$SiO$_2$H]$^+$ were bracketed between two reference measurements of [$^{12}$C$_3$H$_2$F]$^+$ to track variations in the magnetic field strength. The weighted average of the ratio for $\nu_c$([$^{24}$SiO$_2$H]$^+$)/$\nu_c$([$^{12}$C$_3$H$_2$F]$^+$) was determined to be $\bar{R}=1.000085151(71)$ with an associated Birge Ratio \cite{Birge1932} of 1.03. Most systematic uncertainties in $\bar{R}$ scale linearly with the mass difference between the ion of interest and the reference ion \cite{Bollen1990}. This mass-dependent shift has been measured to be $\Delta \bar{R} = 2 \times 10^{-10}$/u~\cite{Gulyuz2015}, which is negligible for isobaric species. Other systematic uncertainties include relativistic effects, ion-ion interactions, and non-linear magnetic field fluctuations. Relativistic effects are negligible compared to the statistical uncertainty \cite{Brown1986}. Ion-ion interactions were limited by excluding shots with more than 5 ions. The non-linear fluctuations in the magnetic field are on the order of $\Delta \bar{R} = 10^{-9}$/hour~\cite{Ringle2007b}. The longest single measurement was 2.5 hours making these fluctuations negligible as well. As a result, the uncertainty of our averaged frequency measurement is statistical.

\begin{figure}
\centering
\includegraphics[width=\columnwidth]{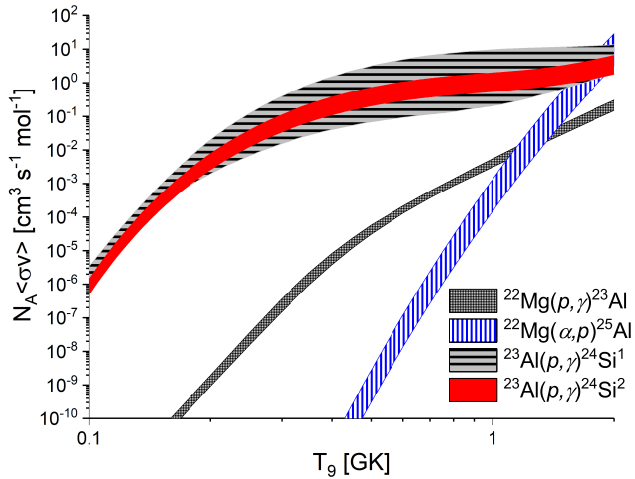}
\caption{The reaction rates for $^{23}$Al($p,\gamma$) (Wolf et al. 2019 ~\cite{Wolf2019}, denoted by the superscript 1, and this work, denoted by the superscript 2), $^{22}$Mg($p,\gamma$) (Iliadis et al. 2010,~\cite{Iliadis2010}) and $^{22}$Mg($\alpha,p$) (Randhawa et al. 2020, \cite{Randhawa2020}) from 0.1 to 2 GK. Our calculations show a dramatic reduction in the 1-$\sigma$ reaction rate uncertainty for $^{23}$Al($p,\gamma$) as a result of our new mass measurement and the corresponding resonance energies.}
\label{fig:ReactionRate}
\end{figure}

The mass excess of $^{24}$Si is extracted directly from $\bar{R}$, and was determined to be $10\;753.8$(37) keV. Our mass is in good agreement with the AME2020 (10 745(19) keV), and a factor of 5 more precise. The AME2020 value is primarily based on a 1980 pion double-charge exchange reaction measurement~\cite{Burleson1980}. With our measurement, a new proton separation energy for $^{24}$Si was calculated to be $3\;283.2$(37) keV. 

With a precisely known $Q$ value, the flow following the $^{22}$Mg waiting point in the $rp$ process is strictly constrained based on nuclear physics information. FIG. \ref{fig:ReactionRate} displays the impact of the new $Q$ value on the uncertainty of the $^{23}$Al($p,\gamma$) reaction rate. Also shown in FIG. \ref{fig:ReactionRate} are the $^{22}$Mg($p,\gamma$) reaction rate, which is represented by the black band, based on information from Ref. \cite{Iliadis2010}, and the reaction rate experimentally determined for the $^{22}$Mg($\alpha, p$) reaction represented by the blue band from Ref. \cite{Randhawa2020}. We estimate the $^{22}{\rm Mg}(\alpha,p)$ reaction rate uncertainty to be $\pm$80\% based on the uncertainty for the lowest-energy cross section measurement from Ref.~\cite{Randhawa2020}.

\begin{table}[htb]
\caption{Resonance levels used in the narrow-resonance approximation for the reaction rate calculation. Experimental information is taken from Ref.~\cite{Wolf2019}. Other levels were calculated with the NuShellX@MSU package with their corresponding spectroscopic factors, radiative widths and proton widths~\cite{Brown2014}, as denoted by *. Excitation energies from NuShellX@MSU calculations were defined with uncertainties of 150 keV~\cite{Henderson2020}.  The reaction rate was calculated with a Monte Carlo program to determine a robust rate uncertainty. A complete description of the reaction rate calculation can be found in the text.}
\begin{ruledtabular}
\begin{tabular}{cccccc}
 $E_x$ (MeV) & $J$ & $l$ & $\text{C}^2\text{S}$ & $\Gamma_\gamma$ (eV) & $\Gamma_p$ (eV) \\
\hline
3.449(5) & 2 & 0 & 0.7(4) & $1.9 \times 10^{-2}$ & $1.0 \times 10^{-4}$ \\
 &  & 2 & 0.002(1) & & \\
 &  & 2 & 0.3(2) & & \\
3.471(6) & 0 & 2 & 0.8(4) & $1.6 \times 10^{-3}$ & $6.2 \times 10^{-5}$ \\
4.256(150)* & 3 & 0 & 0.59 & $1.3 \times 10^{-2}$ & $9.0 \times 10^3$ \\
 &  & 2 & 0.17 &  &  \\
5.353(150)* & 3 & 0 & 0.0012 & $2.8 \times 10^{-2}$ & $3.6 \times 10^3$ \\
 & & 2 & 0.11 & & \\
5.504(150)* & 2 & 0 & 0.044 & $2.2 \times 10^{-1}$ & $2.8 \times 10^4$ \\
 & & 2 & 0.068 &  & \\
5.564(150)* & 4 & 2 & 0.048 & $2.2 \times 10^{-2}$ & $2.2 \times 10^3$ \\
6.004(150)* & 4 & 2 & 0.28 & $6.5 \times 10^{-3}$ & $2.8 \times 10^4$ \\
6.056(150)* & 0 & 2 & 0.053 & $3.4 \times 10^{-3}$ & $5.6 \times 10^3$ \\
6.072(150)* & 2 & 0 & 0.012 & $5.0 \times 10^{-2}$ & $2.4 \times 10^4$ \\
 & & 2 & 0.093 &  & 
\end{tabular}
\end{ruledtabular}
\label{tab:ExcitedStates}
\end{table}

The $^{23}{\rm Al}$($p,\gamma$) reaction rate was calculated using the narrow resonance approximation~\cite{Fowler1967}. The information that contributed to this calculation can be found in Table \ref{tab:ExcitedStates}, based on experimental information~\cite{Wolf2019}, and, for higher lying states, on calculations with NuShellX@MSU~\cite{Brown2014} that were included in the reaction rate calculation of Ref. \cite{Wolf2019}. Based on a previous measurement of longitudinal momentum distributions of $^{24}{\rm Si}$ states~\cite{Longfellow2020}, we assigned the 3471 state with J=0 and used the information from Ref.~\cite{Wolf2019}. 

The reaction rate properties varied using a Monte Carlo random sampling approach. The excitation energies in the unbound states of $^{24}{\rm Si}$ and the $Q$ value for the $^{23}{\rm Al}(p,\gamma)$ reaction were varied, assuming a normal distribution. The impact of the proton penetration factor was also included in the reaction rate calculation~\cite{Humblet1987}. The excitation energies derived from NuShellX@MSU calculations were defined with uncertainties of 150 keV~\cite{Henderson2020}. The radiative widths and spectroscopic factors were varied with a Log-normal distribution~\cite{Longland2010}. All radiative partial widths assumed a factor uncertainty of 2~\cite{Richter2020}. The same factor uncertainty was used for all theoretical spectroscopic factors. Experimental spectroscopic factors and uncertainties were varied using relations found in Ref. \cite{Longland2010}. The proton partial widths were scaled, based on the impact of the spectroscopic factor variation.

For the direct capture component of the reaction rate calculation, the spectroscopic factors for each astrophysical S-factor reported in Ref. \cite{Banu2012} were varied with a Log-normal distribution as well. The experimental ground state spectroscopic factor was varied with its experimental uncertainty while the three theoretical values calculated for the first excited state assumed a factor uncertainty of 2. After varying the spectroscopic factor, the corresponding astrophysical S-factors $S(E_0)$ were scaled to calculate the new S-factor before summing all four contributions in the direct capture component calculation. The results for the calculated rate for the $^{23}{\rm Al}$($p,\gamma$) rate can be found in Table \ref{tab:ReactionRate}.

\begin{table*}[htb]
\caption{The recommended reaction rate based on information from Ref.~\cite{Wolf2019} and this work. Also included are the lower (16th percentile) and upper (64th percentile) uncertainties of this reaction.}
\begin{ruledtabular}
\begin{tabular}{cccc}
 $T_9$ &  & $N_A<\sigma v>$ ($\text{cm}^3/\text{s}/\text{mole}$) &  \\\cmidrule{2-4}
 & Recommended & Lower & Upper \\
 \hline
 0.1 & $8.195 \times 10^{-7}$ & $4.777 \times 10^{-7}$ & $1.402 \times 10^{-6}$\\
 0.15 & $2.754 \times 10^{-4}$ & $1.640 \times 10^{-4}$ & $4.642 \times 10^{-4}$ \\
 0.2 & $4.481 \times 10^{-3}$ & $2.602 \times 10^{-3}$ & $7.761 \times 10^{-3}$ \\
 0.3 & $6.168 \times 10^{-2}$ & $3.432 \times 10^{-2}$ & $1.114 \times 10^{-1}$ \\
 0.4 & $2.022 \times 10^{-1}$ & $1.099 \times 10^{-1}$ & $3.746 \times 10^{-1}$ \\
 0.5 & $3.834 \times 10^{-1}$ & $2.052 \times 10^{-1}$ & $7.212 \times 10^{-1}$ \\
 0.6 & $5.594 \times 10^{-1}$ & $2.967 \times 10^{-1}$ & 1.064 \\
 0.7 & $7.096 \times 10^{-1}$ & $3.745 \times 10^{-1}$ & 1.360 \\
 0.8 & $8.328 \times 10^{-1}$ & $4.392 \times 10^{-1}$ & 1.601 \\
 0.9 & $9.381 \times 10^{-1}$ & $4.950 \times 10^{-1}$ & 1.801 \\
 1.0 & 1.034 & $5.482 \times 10^{-1}$ & 1.979 \\
 1.5 & 1.707 & $9.329 \times 10^{-1}$ & 3.303 \\
 2.0 & 3.222 & 2.010 & 6.553
\end{tabular}
\end{ruledtabular}
\label{tab:ReactionRate}
\end{table*}

\begin{figure}
\begin{center}
\includegraphics[width=\columnwidth]{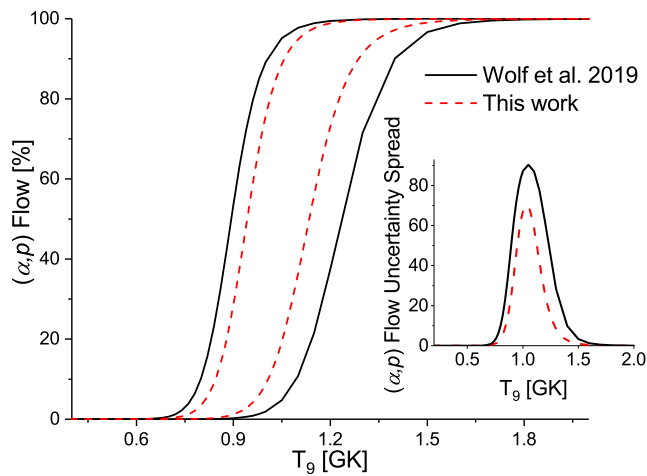}
\end{center}
\caption{Flow through $^{22}$Mg($\alpha,p$) for the temperature range of relevance for X-ray bursts, where the solid black lines assume the $^{23}{\rm Al}(p,\gamma)$ reaction rate uncertainty of Ref.~\cite{Wolf2019} and the dashed red lines correspond to the uncertainty from this work. The inset shows the uncertainty spread in the ($\alpha,p$) flow, which peaks at temperatures reached during the light curve rise in X-ray burst model calculations.}
\label{fig:flow}
\end{figure}

We assessed the impact of the new $^{24}{\rm Si}$ mass on the onset of the $\alpha p$-process by calculating the flow at the $^{22}{\rm Mg}$ bottleneck. To first order, at this point in the reaction network, there is a competition between $^{23}{\rm Al}(p,\gamma)$ and $^{22}{\rm Mg}(\alpha,p)$, given the rather long half-life of $^{22}{\rm Mg}$, 3.88~s~\cite{Sham2015}, and low proton separation energy $S_{p}$ of $^{23}{\rm Al}$, ~0.141~MeV~\cite{Huan2021}. 
The flow into the $\alpha p$-process is defined as $\lambda_{\alpha,p}/\left(\lambda_{p,\gamma}+\lambda_{\alpha,p}\right)$. The rates are $\lambda_{i}=W\left(X_{i}/A_{i}\right)\rho N_{\rm A}\langle\sigma v\rangle$, where $X_{i}$ and $A_{i}$ are the mass fraction and mass number of hydrogen and helium for the $^{23}{\rm Al}(p,\gamma)$ and $^{22}{\rm Mg}(\alpha,p)$ reactions, respectively~\cite{Merz2021}. $W$ is a weight factor, which is the equilibrium abundance of the nuclide as determined by the Saha equation~\cite{Meisel2018,Brown15}. Since the $^{22}{\rm Mg}$ $\beta^{+}$-decay rate is too slow to compete, only strong and electromagnetic interactions are involved in the flow, and the density dependence cancels. For $X_{i}$, we take the conditions determined by Ref.~\cite{Fisker2008} for the X-ray burst ignition region, where $X_{\rm H}=0.03$ and $X_{\rm He}=0.31$.

Results are shown in FIG.~\ref{fig:flow} over the range of temperatures of relevance for X-ray bursts. For the following discussion, we adopt 10\% of flow through $^{22}{\rm Mg}(\alpha,p)$ as the point when the $\alpha p$-process becomes significant. Previously, using the $^{23}{\rm Al}(p,\gamma)$ reaction rate uncertainty from Ref.~\cite{Wolf2019}, the uncertainty band for when the $\alpha p$-process becomes significant is $T_{9}=0.94\pm0.15$~GK. With the new $^{24}{\rm Si}$ mass, this uncertainty band becomes $0.93\pm0.08$~GK, precisely pinpointing the onset of the $\alpha p$-process in X-ray bursts. While adopting other ignition conditions shifts the temperature for significant $\alpha p$-process flow, e.g. $0.99\pm0.09$~GK for the $X_{\rm H}=0.06$ $X_{\rm He}=0.19$ ignition conditions from the calculations of Ref.~\cite{Merz2021}, the precision remains 9\%. Therefore, we firmly establish the relevance of the $\alpha p$-process in powering X-ray burst light curves by determining under which conditions there is $(\alpha,p)$ breakout from the $^{22}{\rm Mg}$ waiting-point.

This Letter presents the first high-precision mass measurement of $^{24}{\rm Si}$ using Penning trap mass spectrometry, resulting in a mass excess of $10\;753.8$(37)~keV. The achieved precision reduces the uncertainty of the $^{23}{\rm Al}(p,\gamma)$ reaction rate, pinpointing the temperature when the the $\alpha p$-process turns-on in X-ray bursts with $^{22}{\rm Mg}(\alpha,p)$ to 9\%. Further refinement would require more precise resonance strength determinations for $^{23}{\rm Al}(p,\gamma)$, as was recently submitted for publication~\cite{Lota21}, and a measurement of the $^{22}{\rm Mg}(\alpha,p)$ cross section to lower center-of-mass energies. 

\begin{acknowledgments}
This material is based upon work supported by the U.S. National Science Foundation through Grants No. PHY-1565546, PHY-2111185 (The Precision Frontier at FRIB: Masses, Radii, Moments, and Fundamental Interactions), PHY-1430152 (Joint Institute for Nuclear Astrophysics -- Center for the Evolution of the Elements), and No. PHY-191355, as well as the U.S. Department of Energy, Office of Science, Office of Nuclear Physics under Grants No. DE-FG02-88ER40387, DE-SC0019042, DE-NA0003909, and DE-SC0015927.  A.A.V. acknowledges support from NSERC (Canada) under Contract No. SAPPJ2018-00028. Part of this work was performed under the auspices of the U.S. Department of Energy by Lawrence Livermore National Laboratory under Contract DE-AC52-07NA27344.
\end{acknowledgments}


\bibliographystyle{apsrev4-2}
\bibliography{24Si.bib}

\end{document}